\documentclass[12pt]{article}
\usepackage{amsmath}
\usepackage{graphicx}
\usepackage{enumerate}
\usepackage{natbib}
\usepackage{url} 

\usepackage{amssymb,physics}
\usepackage{mathrsfs,bm}
\usepackage[svgnames]{xcolor}
\usepackage{enumitem}
\usepackage{multirow,multicol}
\usepackage[ruled]{algorithm2e}

\usepackage{amsthm}
\newtheorem{theorem}{Theorem}

\newtheorem{proposition}{Proposition}

\theoremstyle{remark}

\newtheorem{example}{Example}
\theoremstyle{definition}

\newtheorem{assumption}{Assumption}

\usepackage{xr-hyper}
\usepackage[bookmarksnumbered=true, colorlinks,
	citecolor=ForestGreen,       
	linkcolor=Blue,		         
	urlcolor=Magenta,	     	 
]{hyperref}
\makeatletter
\newcommand*{\addFileDependency}[1]{
\typeout{(#1)}
\@addtofilelist{#1}
\IfFileExists{#1}{}{\typeout{No file #1.}}
}
\makeatother
\newcommand*{\myexternaldocument}[1]{%
\externaldocument{#1}%
\addFileDependency{#1.tex}%
\addFileDependency{#1.aux}%
}
\myexternaldocument{supp-JASA}

\graphicspath{{./art/}}

\renewcommand{\Pr}{\mathbb{P}}
\newcommand{\E}{\mathbb{E}}

\begin{document}

\def\spacingset#1{\renewcommand{\baselinestretch}%
{#1}\small\normalsize} \spacingset{1}

\newcommand{\blind}{0}

\if0\blind
{
  \title{\bf Function-on-function Differential Regression}
  \author{Tongyu Li and Fang Yao\thanks{
    Fang Yao is the corresponding author: \texttt{fyao@math.pku.edu.cn}. This research is partially supported by  the National Key R\&D Program of China (No. 2022YFA1003801), the National Natural Science Foundation of China (No. 12292981), the Newcorner Stone Science foundation through the Xplorer Prize, the LMAM and the Fundamental Research Funds for the Central Universities, Peking University (LMEQF).}\\
    Department of Probability \& Statistics, School of Mathematical Sciences,\\ Center for Statistical Science, Peking University}
    \date{}
  \maketitle
} \fi

\if1\blind
{
  \bigskip
  \bigskip
  \bigskip
  \begin{center}
    {\LARGE\bf Function-on-function Differential Regression}
\end{center}
  \medskip
} \fi

\bigskip
\begin{abstract}
Function-on-function regression has been a topic of substantial interest due to its broad applicability, where the relation between functional predictor and response is concerned. In this article, we propose a new framework for modeling the regression mapping that extends beyond integral type, motivated by the prevalence of physical phenomena governed by differential relations, which is referred to as function-on-function differential regression. However, a key challenge lies in representing the differential regression operator, unlike functions that can be expressed by expansions. As a main contribution, we introduce a new notion of model identification involving differential operators, defined through their action on functions. Based on this action-aware identification, we are able to develop a regularization method for estimation using operator reproducing kernel Hilbert spaces. Then a goodness-of-fit test is constructed, which facilitates model checking for differential regression relations. We establish a Bahadur representation for the regression estimator with various theoretical implications, such as the minimax optimality of the proposed estimator, and the validity and consistency of the proposed test. To illustrate the effectiveness of our method, we conduct simulation studies and an application to a real data example on the thermodynamic energy equation.
\end{abstract}

\noindent%
\vfill

\newpage
\spacingset{1.9} 

\section{Introduction}
\label{sec:intro}
\subsection{Literature review and motivation}

Due to technological advances in data collection, there has been a rising interest in analysis of functional data that are inherently functions, curves or surfaces over a continuum like time and spectrum; see \citet{ramsay2005functional,wang2016functional} for an introduction to the field of functional data analysis (FDA). 
A large body of literature extends statistical methods to handle functional data, offering new insights into the complex relationships embedded in many real-world applications, such as in medicine \citep{sorensen2013medical}, health science \citep{acar2024wearsensor}, biomechanics \citep{gunning2024biomechanics} and computational biology \citep{cremona2019biology}. 
One of the most important areas within FDA is functional regression, where the relation between functional predictor and response is modeled using various structures and may be complicated by data designs and sampling schemes. 
The extensive studies on function-on-function regression models include \citet{yao2005functional,sun2018optimal,luo2017function,zhang2013time,kim2018additive,jiang2011functional,lian2007nonlinear,luo2024general,rao2023modern,kovachki2023neural,reinhardt2024statistical}, to name a few. 
These functional models are analogous to their multivariate counterparts, allowing the adaptation of existing multivariate methods \citep{greven2017general}, for which the prevalent line of research has focused on integral-type relations, i.e., the regression mapping is identified with some coefficient function by means of a certain integral transformation. 

However, the functional form of the relationship between response and predictor often incorporates derivative information, since analytical laws governing physical phenomena are typically expressed through differential equations \citep{ibragimov2009practical,chicone2016invitation}. 
Differential equations are mathematical tools that describe how a quantity changes with respect to one or more variables, capturing the dynamics of systems in physics, engineering, biology, and other fields. 
Taking such derivatives into account necessitates new regression models for functional data, which can lead to more accurate prediction. 
It is also noteworthy that statistical practice has found (partial) derivative data to be beneficial in improving the efficiency of nonparametric estimation \citep{hall2007nonparametric,hall2010nonparametric,dai2024nonparametric}, but the procedures for modeling differential relation between functions are still absent. 

A distinct yet related topic is dynamic data analysis \citep{ramsay2017dynamic} that leverages statistical learning in differential dynamics to facilitate data-driven methods for recovering the solution path and inferring the structure of dynamical systems. By specifying a parametric family of differential operators, \citet{ramsay2007parameter,xun2013parameter,brunel2014parametric,yang2021inference,tan2024green} addressed trajectory reconstruction and uncertainty quantification using flexible parameter estimation techniques. 
The parametric approach has been extended to handle mixed effects \citep{cao2010linear,lu2011high,wang2014estimating} and varying coefficients \citep{xue2010sieve,zhang2017estimating}. 
Recent years have witnessed growing attention to nonparametric modeling of differential equations, whose learning procedures are termed as equation discovery \citep{bongard2007automated,schmidt2009distilling}. 
The considered differential operators are constructed from a predetermined dictionary of candidate differentiation features, e.g., $\mathcal{D}(u) = f(u, \partial_{t}u, \nabla u, \nabla^{2} u)$ with some unknown function $f$. 
This framework, usually referred to as symbolic regression, has been utilized by \citet{brunton2016discovering,rudy2017data,raissi2018deep,long2019pde,long2024kernel} but still lacks theoretical guarantees. 

Although there are many methods for estimation and inference for a certain differential equation in dynamic data analysis, the primary focus is on a single solution trajectory $u(\cdot)$. The typical observations therein are noisy measurements of $u(x_i)$ for some discrete points $x_i$, $i=1,\dots,n$. 
By contrast, function-on-function regression aims at modeling  realizations of a pair of random functions, which we denote by $(U_i,F_i)$, $i=1,\dots,n$. 
We are interested in the regression mapping between $U_i$ and $F_i$ that incorporates differential relations, and such a regression problem has not been addressed in literature. 
Consequently, we need to develop new methods to deal with such a function-on-function differential regression model.

\subsection{Formulation of differential regression and examples}
In this work, we tackle regression analysis for differential operators that link functional responses to functional predictors. The assumed model is 
\begin{equation}\label{eq:model}
	F = \mathcal{D}(U) + \varepsilon ,
\end{equation}
where $F$ is a functional response, $U$ is a functional predictor, $\varepsilon$ is a zero-mean error independent of $U$, and $\mathcal{D}$ is an unknown differential operator, which possibly belongs to some expert-designed family grounded in mathematical and physical intuition. 
Several noiseless examples, from the wide range of continuum physics \citep{morro2023mathematical}, are shown below. 
\begin{enumerate}
	\item\emph{Thermal conduction}. 
	Heat energy is transferred from a higher temperature area to a lower one, which abides by Fourier's law \[ q = - \kappa \nabla T .\] Here the response $q$ is the local heat flux density, the predictor $T$ is the temperature, and the differential operator $- \kappa \nabla$ is a constant multiple of negative gradient, with $\kappa$ being the conductivity of the considered material.
	\item\emph{Electromagnetic induction}. 
	Faraday's law indicates how a time-varying magnetic field would interact with an electric circuit to produce an electromotive force, formally given by \[ E = - N \partial_{t} \Phi ,\] where the response $E$ is the induced electromotive force, the predictor $\Phi$ is the magnetic flux, and the differential operator $- N \partial_{t}$ denotes the negative time derivative multiplied by the number of turns.
	\item\emph{Fluid mechanics}. 
	The motion of viscous fluid substances is described by the celebrated Navier--Stokes equations. For incompressible fluids, the convective form is \[ \partial_{t}u + u \bm{\cdot} \nabla u - \nu \nabla^{2} u = - (1/\rho) \nabla p + g ,\] where $u$ is the flow velocity, $\rho$ is the mass density, $p$ is the pressure, $g$ represents body force, $\nu$ is the dynamic viscosity, $\partial_t$ is the time derivative and $\nabla$ denotes the del operator with respect to spatial variables. While the left-hand side can be viewed as the functional predictor $u$ acted upon by a differential operator, the right-hand side as a whole corresponds to the response.
\end{enumerate}

\subsection{Our contributions}

Instead of deriving the differential operator $\mathcal{D}$ from first principles, we try to learn from data generated by the model \eqref{eq:model}. 
To conduct nonparametric analysis for the regression model \eqref{eq:model}, we first propose an action-aware identification of the differential operator $\mathcal{D}$, which implicitly accounts for the underlying smoothness by inspecting the action of $\mathcal{D}$ on functions. 
Different from the existing approaches, the differential operator under consideration is not limited to a specified differentiation dictionary, and thus favors more generality at the level of preserving all information of possible derivatives. 
We then construct a regularization estimator in an operator reproducing kernel Hilbert space framework \citep{kadri2016operator,stepaniants2023learning}, which admits an easy-to-implement form by an operator representer theorem. 
This estimation procedure enables us to handle various inference problems. 
In particular, we develop a goodness-of-fit test based on a linear smoother regarding the nonparametric fitting, which helps in verifying an expert-designed differential operator. 

The main contributions of this paper are summarized below. 
First, to our best knowledge, this is the first investigation of statistical inference for the function-on-function differential model \eqref{eq:model}. In contrast to the commonly used methods of identification, we propose an action-aware approach that is more general in the nonparametric setting. 
Second, we provide a regularization method for estimation, based on which a Bahadur representation is derived to serve as the foundation of nonparametric inference. 
Third, we develop a goodness-of-fit test for the differential operator, together with a bootstrap procedure tailored to provide critical values. 
Finally, the regression estimator is proved to achieve minimax optimality, and the validity and consistency of the proposed test are established. 
The theoretical findings are supported by simulation studies and an application to a real data example on the thermodynamic energy equation. 

The rest of this paper is organized as follows. 
Section~\ref{sec:method} details our estimation and testing procedures for the inference about \eqref{eq:model}, and presents a representer theorem that simplifies the implementation. 
In Section~\ref{sec:theoretical}, we examine the theoretical results around the proposed method. 
The favorable numerical performance is illustrated by experiments in Sections \ref{sec:simulation} and \ref{sec:realdata}. 
All technical proofs and auxiliary lemmas, as well as the code for producing numerical results, are collected in the Supplementary Material.

\section{Proposed Methodology}
\label{sec:method}
For the sake of clarity, following \citet{horvath2012inference} among others, we focus on the case where functional data are fully observed, which typically originate from pre-smoothing of measurements on dense grids. 
Suppose that the observations $(U_{i},F_{i})$, $i=1,\dots,n$, are independent copies of $(U,F)$ in the functional regression model \eqref{eq:model}. 
We are interested in recovering the differential operator $\mathcal{D}$, and investigating related inference problems. 
In what follows, we begin by introducing an action-aware identification method, which facilitates the statistical analysis in a general nonparametric setting. 

\subsection{Action-aware identification of differential operator}
Due to the difficulty of viewing them as elements in a linear space, it is challenging to represent differential operators in a straightforward way, unlike functions that can be written as basis expansions. 
Although one may construct a differential operator using some predetermined differentiation features, possible omissions of derivative information could obstacle subsequent estimation. 
To achieve more flexibility and generality, we identify a differential operator via its action on functions. 
As a starting point, the following Example~\ref{ex:action} is intended for a more concrete understanding. 
\begin{example}\label{ex:action}
	Consider the differential operator $\mathcal{D} = \nabla^{2} + \omega^{2}$ acting on $\mathcal{C}^{2}([0,1])$, where $\nabla^{2}$ is the Laplacian operator, $\omega$ is a nonnegative number, and $\mathcal{C}^{2}([0,1])$ consists of twice differentiable functions on the interval $[0,1]$. 
	For any $u \in \mathcal{C}^{2}([0,1])$, simple calculations give 
	\[ u(y) = u(0) + y \nabla u(0) + \int_{0}^{1} \Phi(x,y) \nabla^{2}u(x) \dd{x} ,\]
	\[ \mathcal{D}u(y) = \omega^{2} u(0) + \omega^{2} y \nabla u(0) + \int_{0}^{1} \{\delta(y-x) + \omega^{2}\Phi(x,y)\} \nabla^{2}u(x) \dd{x} ,\]
	where $\Phi(x,y) = \max(y-x,0)$ and $\delta(\cdot)$ is the Dirac delta function. 
	Let $\tilde{u} = (u(0), \nabla u(0), \nabla^{2}u)$. Then $u$ corresponds uniquely to $\tilde{u}$, and thus $\mathcal{D}$ is determined by the generalized function $(\omega^{2}, \omega^{2}y, \delta(y-x) + \omega^{2}\Phi(x,y))$. 
	Since $\delta(y-x) + \omega^{2}\Phi(x,y) = \nabla^{2}_{y}\{\Phi(x,y) + \omega^{2}\Phi^{*2}(x,y)\}$, where $\Phi^{*2}(x,y) = \int_{0}^{1} \Phi(x,z) \Phi(z,y) \dd{z}$, the regularity of the differential operator $\mathcal{D}$ could be attributed to the integrability of the function $(x,y) \in [0,1]^{2} \mapsto \Phi(x,y) + \omega^{2}\Phi^{*2}(x,y) \in \mathbb{R}$. 
\end{example}

Now we present a general framework for identification of differential operators. 
Denote by $\Omega$ the domain on which \eqref{eq:model} holds, and let $\Gamma$ be the boundary of $\Omega$. 
Since functions $u$ in the domain of the differential operator $\mathcal{D}$ ought to possess certain regularity, we assume them to be solutions of a partial differential equation 
\begin{equation}\label{eq:pde}
	\mathcal{P}(u) = f \text{ in } \Omega ,\quad \mathcal{B}(u) = g \text{ on } \Gamma ,
\end{equation}
where $\mathcal{P},\mathcal{B}$ are predetermined differential operators, and $f,g$ are forcing and source functions that vary to produce different solutions $u$. 
Then we can write $u = \mathcal{A}(\tilde{u})$, where $\mathcal{A}$ is the solution operator for \eqref{eq:pde}, and $\tilde{u}$ is a shorthand for $(\mathcal{P}(u),\mathcal{B}(u))$. 
Clearly \[ \mathcal{D}(u) = \mathcal{D}\circ\mathcal{A}(\tilde{u}) ,\] so $\mathcal{D}$ corresponds uniquely to $\mathcal{D}\circ\mathcal{A}$, which utilizes the fact that $\mathcal{A}\circ(\mathcal{P},\mathcal{B})$ is the identity map. 
This transformation of $\mathcal{D}$ yields better tractability due to integration within $\mathcal{A}$. 

Before proceeding, we exemplify a class of solution operators that are easy to handle. 
The analysis of general solution operators has undergone extensive studies in its own right \citep{boulle2024mathematical}.
\begin{example}[Linear solution operator]\label{ex:Green}
	If $\mathcal{P}$ is a linear partial differential operator and $\mathcal{B}$ is a linear boundary condition, then the solution $u$ to \eqref{eq:pde} admits the following representation \citep[Chapter 2, Section 2]{lions2012non}:
	\[ u = \mathcal{A}(f,g) = \int_{\Omega} G(x,\bm{\cdot}) f(x) \dd{x} + \int_{\Gamma} G_{\partial}(x,\bm{\cdot}) g(x) \dd{S(x)} ,\]
	where $G$ is called the Green's function \citep[Section~8.7]{arbogast2025functional}, $G_{\partial}$ is a modification of $G$ using $\mathcal{B}$, and $\dd{S}$ denotes the surface measure on $\Gamma$. 
	The solution operator $\mathcal{A}$ is determined by $(G,G_{\partial})$, which corresponds uniquely to $(\mathcal{P},\mathcal{B})$. 
\end{example}

In order to apply the rich arsenal for dealing with integral-type operators, we further import a linear differential operator $\mathcal{L}$ and let $\mathcal{D}$ correspond uniquely to the operator 
\begin{equation}\label{eq:operator}
	\mathcal{T} = \mathcal{L}^{-1}\circ\mathcal{D}\circ\mathcal{A} ,
\end{equation}
where $\mathcal{L}^{-1}$ is the solution operator associated with the partial differential equation 
\[ \mathcal{L}u = f \text{ in } \Omega ,\quad u = 0 \text{ on } \Gamma .\]
The reduction to \eqref{eq:operator} is justified by the observation that $\mathcal{L}\circ\mathcal{T} = \mathcal{D}\circ\mathcal{A}$, since $\mathcal{L}\circ\mathcal{L}^{-1}$ turns out to be the identity map. 
Now the model \eqref{eq:model} is transformed into 
\begin{equation}\label{eq:model_trans}
	F = \mathcal{D}\circ\mathcal{A}(\tilde{U}) + \varepsilon = \mathcal{L}\mathcal{T}(\tilde{U}) + \varepsilon ,
\end{equation}
and it suffices to perform inference on the operator $\mathcal{T}$. 
Once an estimator $\hat{\mathcal{T}}$ for $\mathcal{T}$ is available, we shall estimate $\mathcal{D}$ by $\hat{\mathcal{D}}$ such that $\hat{\mathcal{D}}(u) = \mathcal{L}\hat{\mathcal{T}}(\tilde{u})$. 
Note that the linearity of $\mathcal{L}$ is essential in that $\E(F\mid U)$ depends linearly on the unknown $\mathcal{T}$.

\subsection{Regularized estimation and representer theorem}
Taking into account the smoothness within the model \eqref{eq:model_trans}, we assume that $\mathcal{T}$ resides in an operator reproducing kernel Hilbert space $\mathcal{H}$, a subcollection of operators from $\mathcal{V} = L^2(\Omega) \oplus L^2(\Gamma)$ to $L^2(\Omega)$, where $L^2(\Omega)$ denotes the space of square integrable functions on $\Omega$. 
See \cite{kadri2016operator} for preliminaries. 
One may choose $\mathcal{H}$ to be some class of solution operators, e.g., integral operators associated with Green's functions as Example~\ref{ex:Green}. 
\begin{example}\label{ex:opRKHS}
	Let $\mathcal{H} = \{\mathcal{O}_{G,G_\partial}: G\in\mathcal{H}_{\Omega} ,\, G_{\partial}\in\mathcal{H}_{\Gamma}\}$, where $\mathcal{O}_{G,G_\partial}$ is defined as \[ \mathcal{O}_{G,G_\partial} : (f,g) \mapsto \int_{\Omega} G(x,\bm{\cdot}) f(x) \dd{x} + \int_{\Gamma} G_{\partial}(x,\bm{\cdot}) g(x) \dd{S(x)} ,\] and $\mathcal{H}_{\Omega}\subset L^2(\Omega\times\Omega)$ and $\mathcal{H}_{\Gamma}\subset L^2(\Gamma\times\Omega)$ are reproducing kernel Hilbert spaces. 
	This is motivated by Example~\ref{ex:action} where $\mathcal{T} = \mathcal{O}_{G,G_\partial}$ with $G(x,y) = \Phi(x,y) + \omega^{2}\Phi^{*2}(x,y)$. 
	Denote the reproducing kernels of $\mathcal{H}_{\Omega}$ and $\mathcal{H}_{\Gamma}$ by $K : (\Omega\times\Omega)^{2} \to \mathbb{R}$ and $K_{\partial} : (\Gamma\times\Omega)^{2} \to \mathbb{R}$, respectively. 
	If $\mathcal{H}$ is endowed with the inner product \[ \langle\mathcal{O}_{G^{(1)},G^{(1)}_\partial},\mathcal{O}_{G^{(2)},G^{(2)}_\partial}\rangle_{\mathcal{H}} = \langle G^{(1)},G^{(2)} \rangle_{\mathcal{H}_\Omega} + \langle G^{(1)}_{\partial}, G^{(2)}_{\partial} \rangle_{\mathcal{H}_\Gamma} ,\] then it follows from \citet[Remark 12]{stepaniants2023learning} that $\mathcal{H}$ becomes an operator reproducing kernel Hilbert space with the operator-valued reproducing kernel $\mathcal{K}$ given by 
	\[\begin{aligned}
		\mathcal{K}(\tilde{u},\tilde{v})f &= \int_{\Omega^3} K(x,\bm{\cdot},\xi,\eta) \mathcal{P}(v)(x) \mathcal{P}(u)(\xi) f(\eta) \dd{x}\dd{\xi}\dd{\eta} \\ &\quad + \int_{\Gamma^{2}\times\Omega} K_{\partial}(x,\bm{\cdot},\xi,\eta) \mathcal{B}(v)(x) \mathcal{B}(u)(\xi) f(\eta) \dd{S(x)}\dd{S(\xi)}\dd{\eta} .
	\end{aligned}\]
\end{example}

We propose a regularization estimator $\hat{\mathcal{T}}_{\lambda}$ for $\mathcal{T}$ in \eqref{eq:model_trans}, obtained by minimizing $P_{n}\ell_{\mathcal{O};\lambda} = n^{-1}\sum_{i=1}^{n}\ell_{\mathcal{O};\lambda}(\tilde{U}_{i},F_{i})$ among $\mathcal{O}\in\mathcal{H}$ for a tuning parameter $\lambda>0$, i.e.,
\begin{equation}\label{eq:est}
	\hat{\mathcal{T}}_{\lambda} = \operatorname{arg\,min}_{\mathcal{O}\in\mathcal{H}} P_{n}\ell_{\mathcal{O};\lambda} ,
\end{equation}
where $P_{n}$ is the empirical measure and $\ell_{\mathcal{O};\lambda}$ is the regularized loss function defined by 
\begin{equation}\label{eq:loss}
	\ell_{\mathcal{O};\lambda}(\tilde{u},f) = \norm{f-\mathcal{L}\mathcal{O}(\tilde{u})}_{L^2(\Omega)}^{2} + \lambda \norm{\mathcal{O}}_{\mathcal{H}}^{2} ,\quad \tilde{u}\in\mathcal{V}, \, f\in L^2(\Omega) .
\end{equation}
Here $\norm{\bm{\cdot}}_{L^2(\Omega)}$ (resp.\ $\norm{\bm{\cdot}}_{\mathcal{H}}$) is the norm induced by the inner product $\langle\bm{\cdot},\bm{\cdot}\rangle_{L^2(\Omega)}$ (resp.\ $\langle\bm{\cdot},\bm{\cdot}\rangle_{\mathcal{H}}$) of $L^2(\Omega)$ (resp.\ $\mathcal{H}$). 
The function $\ell_{\mathcal{O};\lambda}$ consists of two parts assessing the fidelity to the data and the plausibility of the model, respectively, which is balanced via $\lambda$. 

Next, we show that the minimization problem is easily computed in a finite-dimensional subspace of $\mathcal{H}$. 
Let $\mathcal{K}$ be the operator-valued reproducing kernel associated with $\mathcal{H}$, which is a mapping from $\mathcal{V}\times\mathcal{V}$ to the space $\mathbb{B}(L^2(\Omega))$ of bounded linear operators on $L^2(\Omega)$. 
Thanks to the reproducing property of $\mathcal{K}$, we have the following representation for $\hat{\mathcal{T}}_{\lambda}$. 
\begin{proposition}[Representer theorem]\label{prop:rep}
	The minimizer \eqref{eq:est} exists and is given by 
	\[ \hat{\mathcal{T}}_{\lambda} = \sum_{i=1}^{n} \mathcal{K}(\tilde{U}_{i},\bm{\cdot})h_{i} \quad\text{for some } h_{1},\dots,h_{n} \in L^2(\Omega) ,\]
	and moreover, the vector $(h_{i}) \in (L^2(\Omega))^n$ is a linear transformation of $(F_{i}) \in (L^2(\Omega))^n$. 
	In particular, if $\mathcal{H}$ only contains linear operators, $(\mathcal{P},\mathcal{B}) : u \mapsto \tilde{u}$ is a linear operator, and all the functions $U_{i},F_{i},h_{i}$ lie in the linear span of an orthonormal system $\{\phi_{1},\dots,\phi_{p}\} \subset L^2(\Omega)$, then $\hat{\mathcal{T}}_{\lambda} = \sum_{j,k} \hat{c}_{jk} \mathcal{K}(\tilde{\phi}_{k},\bm{\cdot})\phi_{j}$ with $\hat{\mathbf{c}} = (\hat{c}_{jk}) \in \mathbb{R}^{p^2}$ given by 
	\begin{equation}\label{eq:est_linear}
		\hat{\mathbf{c}} = \{\mathbf{K}_{\mathcal{L}}^{\top}(\mathbf{I}_{p}\otimes\mathbf{U}^{\top}\mathbf{U})\mathbf{K}_{\mathcal{L}}^{\top} + n \lambda \mathbf{K}\}^{-1} \{(\mathbf{I}_{p}\otimes\mathbf{U})\mathbf{K}_{\mathcal{L}}\}^{\top}\operatorname{vec}(\mathbf{F}) ,
	\end{equation}
	where $\mathbf{U},\mathbf{F} \in \mathbb{R}^{n \times p}$ are the matrices with entries $\langle U_{i},\phi_{k} \rangle_{L^2(\Omega)}$, $\langle F_{i},\phi_{j} \rangle_{L^2(\Omega)}$, respectively, 
	$\mathbf{K},\mathbf{K}_{\mathcal{L}} \in \mathbb{R}^{p^2 \times p^2}$ are the matrices with entries $\langle\mathcal{K}(\tilde{\phi}_{k},\tilde{\phi}_{k'})\phi_{j},\phi_{j'}\rangle_{L^2(\Omega)}$, $\langle\mathcal{L}\mathcal{K}(\tilde{\phi}_{k},\tilde{\phi}_{k'})\phi_{j},\phi_{j'}\rangle_{L^2(\Omega)}$, respectively, 
	$\mathbf{I}_{p} \in \mathbb{R}^{p\times p}$ is the identity matrix, 
	$\otimes$ denotes the Kronecker product, and $\operatorname{vec}$ denotes the vectorization. 
\end{proposition}

\subsection{Smoothing-based goodness-of-fit test}
In the function-on-function model \eqref{eq:model}, consider testing the goodness of fit for a given family of differential operators $\{\mathcal{D}_{\theta}:\theta\in\Theta\}$, where the parameter space $\Theta$ is an Euclidean subset. 
Denoting $\varepsilon_{\theta} = F - \mathcal{D}_{\theta}(U)$, the hypotheses are formally 
\[\begin{aligned}
	& H_{0} : \E(\varepsilon_{\theta}\mid U) = 0 \ \text{for some } \theta\in\Theta \\
	\text{vs. }& H_{1} : \Pr\{\E(\varepsilon_{\theta}\mid U) = 0\} < 1 \ \text{for all } \theta\in\Theta .
\end{aligned}\]
Scientific conclusions often rely heavily on the specified form of the differential relation, making it a necessary step to check the model plausibility as above. \cite{hart1997nonparametric,gonzalez2023review} reviewed the area of model checking. 
We propose a smoothing-based test for the regression diagnostics, whose construction is connected with the nonparametric estimation \eqref{eq:est} for the transformed model \eqref{eq:model_trans}. 

Proposition~\ref{prop:rep} implies that $\hat{\mathcal{D}}_{\lambda}(u) = \mathcal{L}\hat{\mathcal{T}}_{\lambda}(\tilde{u})$ is a linear smoother, formed by a linear transformation of $(F_{i})$. 
It is immediate that there is a linear operator $\mathcal{S}_{\lambda}$ on $(L^2(\Omega))^n$, depending only on $(U_{i})$, such that the vector of fitted functions $(\hat{F}_{i}) = (\hat{\mathcal{D}}_{\lambda}(U_{i}))$ is given by \[ (\hat{F}_{i}) = \mathcal{S}_{\lambda}\{(F_{i})\} .\]
In the case of \eqref{eq:est_linear}, $\mathcal{S}_{\lambda}(\mathbf{F}\bm{\phi}) = \hat{\mathbf{F}}\bm{\phi}$, where $\bm{\phi} = (\phi_{j})$ and $\operatorname{vec}(\hat{\mathbf{F}}) = \mathbf{S}_{\lambda}\operatorname{vec}(\mathbf{F})$ with 
\begin{equation}\label{eq:smooth}
	\mathbf{S}_{\lambda} = \{(\mathbf{I}_{p}\otimes\mathbf{U})\mathbf{K}_{\mathcal{L}}\} \{\mathbf{K}_{\mathcal{L}}^{\top}(\mathbf{I}_{p}\otimes\mathbf{U}^{\top}\mathbf{U})\mathbf{K}_{\mathcal{L}}^{\top} + n \lambda \mathbf{K}\}^{-1} \{(\mathbf{I}_{p}\otimes\mathbf{U})\mathbf{K}_{\mathcal{L}}\}^{\top} .
\end{equation}
The operator $\mathcal{S}_{\lambda}$ corresponds to smoothing with respect to the predictors, which reflects local information of the data. 

Based on the smoothing operator $\mathcal{S}_{\lambda}$, we construct a fit comparison statistic as follows. 
Let $\hat{\theta}$ be an estimator for $\theta_{0} = \operatorname{arg\,min}_{\theta\in\Theta}\E\{\norm{F-\mathcal{D}_{\theta}(U)}_{L^2(\Omega)}^{2}\}$, the true value of $\theta$ under the null model. For example, one could set $\hat{\theta} = \operatorname{arg\,min}_{\theta\in\Theta}\sum_{i=1}^{n}\norm{F_{i}-\mathcal{D}_{\theta}(U_{i})}_{L^2(\Omega)}^{2}$. 
Define $\tilde{\bm{\varepsilon}} = (\tilde{\varepsilon}_{i})$ to be the residuals from the fitted null model, i.e., $\tilde{\varepsilon}_{i} = F_{i}-\mathcal{D}_{\hat{\theta}}(U_{i})$. 
Then we propose the test statistic 
\begin{equation}\label{eq:test}
	Q_{n} = n^{-1} \norm{\mathcal{S}_{\lambda}\tilde{\bm{\varepsilon}}}_{(L^2(\Omega))^n}^{2}
\end{equation}
for $H_0$, which is a sample analog of $\E[\norm{\E\{F - \mathcal{D}_{\theta_0}(U)\mid U\}}_{(L^2(\Omega))^n}^{2}]$. 
It will be shown later that $Q_n$ admits a normal approximation, but with some complicated unknown parameters. 
To provide a practical method of calculating the sampling distribution, especially the critical values for the proposed test, we appeal to the bootstrap \citep{horowitz2019bootstrap}. 

We devise a wild bootstrap procedure inspired by \citet{hardle1993comparing}, which proceeds quickly with repeated generation of multipliers. 
Denote by $B$ the number of resamples. For $b=1,\dots,B$, we generate independent random variables $\delta^{b}_{1},\dots,\delta^{b}_{n}$ identically distributed as $\delta$, which satisfies that $\delta = (1\pm\sqrt{5})/2$ with probability $(\sqrt{5}\mp1)/(2\sqrt{5})$. 
The choice of $\delta$ ensures that $\E(\delta)=0$ and $\E(\delta^{2}) = \E(\delta^{3}) = 1$. 
A data-driven critical value $c_n^B(\alpha)$ is given by the $(1-\alpha)$th sample quantile of 
\begin{equation}\label{eq:btsp}
	Q_{n}^{b} = n^{-1} \norm{\mathcal{S}_{\lambda}\bm{\varepsilon}^{b}}_{(L^2(\Omega))^n}^{2} ,
\end{equation}
where $\bm{\varepsilon}^{b} = (\delta^{b}_{i}\tilde{\varepsilon}_{i})$, or $\bm{\varepsilon}^{b} = (\delta^{b}_{i}\hat{\varepsilon}_{i})$ with $\hat{\varepsilon}_{i} = F_{i} - \hat{\mathcal{D}}_{\lambda}(U_{i})$. 
In other words, the $p$-value of the proposed test is approximated by $1 - B^{-1} |\{b : Q_{n} \ge Q_{n}^{b}\}|$. 

\section{Theoretical Guarantees}
\label{sec:theoretical}
In this section, we conduct error analysis for the proposed regularization method for estimation. 
A core consequence is a Bahadur representation in the sense of \citet{shang2013local}, which serves as an important ingredient for statistical inference. 
In particular, we deduce the minimax optimality of the proposed estimator \eqref{eq:est}, and establish the validity and consistency of the goodness-of-fit test \eqref{eq:test}. 

We first introduce some notation to be used throughout. 
For any nonnegative quantities $A_{\zeta},B_{\zeta}$ depending on $\zeta$, denote $A_{\zeta} \lesssim B_{\zeta}$ if there exists some $C \in \mathbb{R}_{+}$ such that $A_{\zeta} \le C B_{\zeta}$ for all $\zeta$; also write $A_{\zeta} \asymp B_{\zeta}$ if $A_{\zeta} \lesssim B_{\zeta} \lesssim A_{\zeta}$. 
Let $P_n$ and $P$ be the empirical and population measures such that $P_{n}\tau = n^{-1}\sum_{i=1}^{n}\tau(\tilde{U}_{i},F_{i})$ and $P\tau = \E\{\tau(\tilde{U},F)\}$ for any mapping $\tau$ on $\mathcal{V} \times L^{2}(\Omega)$. 
The Orlicz $\psi_{p}$-norm of a random variable $R$ is defined as \citep[Section 2]{vershynin2018high} \[ \norm{R}_{\psi_{p}} = \inf\{\alpha>0 : \E(\mathrm{e}^{|R/\alpha|^{p}}) \le 2\} .\]
Regarding the regression model \eqref{eq:model_trans}, several assumptions on the distribution of $(U,\varepsilon)$ are stated below for 
developing the asymptotic theory. 
\begin{assumption}\label{asm:err}
	The error $\varepsilon$ satisfies that $\big\lVert\norm{\varepsilon}_{L^2(\Omega)}\big\rVert_{\psi_1} < \infty$. 
	Moreover, there exists a constant $C_{\varepsilon} > 0$ such that $C_{\varepsilon}^{-1}\norm{f}_{L^2(\Omega)}^{2} \le \E(\langle\varepsilon,f\rangle_{L^2(\Omega)}^{2}) \le C_{\varepsilon}\norm{f}_{L^2(\Omega)}^{2}$ for any $f \in L^2(\Omega)$. 
\end{assumption}
\begin{assumption}\label{asm:cov}
	The quantity $\norm{\mathcal{O}}_{\Sigma}^{2} = \E\{\lVert\mathcal{L}\mathcal{O}(\tilde{U})\rVert_{L^2(\Omega)}^{2}\}$ is finite for any $\mathcal{O}\in\mathcal{H}$. 
	Let \[ \langle\mathcal{O}_{1},\mathcal{O}_{2}\rangle_{\Sigma} = \E\{\langle\mathcal{L}\mathcal{O}_{1}(\tilde{U}),\mathcal{L}\mathcal{O}_{2}(\tilde{U})\rangle_{L^2(\Omega)}\} ,\quad \mathcal{O}_{1},\mathcal{O}_{2}\in\mathcal{H} .\]
	The operator $\varSigma$ on $\mathcal{H}$ defined by $\langle\mathcal{O}_{1},\mathcal{O}_{2}\rangle_{\Sigma} = \langle\varSigma\mathcal{O}_{1},\mathcal{O}_{2}\rangle_{\mathcal{H}}$ is compact. 
	Furthermore, there exists a constant $C_{U} > 0$ such that $\big\lVert\lVert\mathcal{L}\mathcal{O}(\tilde{U})\rVert_{L^2(\Omega)}\big\rVert_{\psi_1} \le C_{U} \norm{\mathcal{O}}_{\Sigma}$ for any $\mathcal{O}\in\mathcal{H}$. 
\end{assumption}
Assumptions~\ref{asm:err}--\ref{asm:cov} are commonly adopted in the context of functional regression \citep[confer, e.g.,][]{dette2024statistical}. The upper bounds on the exponential moments ensure that the tail probabilities of the quantities of interest in the model are controllable. Besides, our characterization of the covariance structure of the noise and covariates gives a hint of the behavior of the estimator based on least squares. 
In Assumption~\ref{asm:err}, a prototype of $\varepsilon$ is the Gaussian white noise error process. 
To help understanding Assumption~\ref{asm:cov}, we provide the following example. 
\begin{example}\label{ex:Cov}
	Consider $\mathcal{O}_{G,G_\partial}$ defined in Example~\ref{ex:opRKHS}. 
	By Minkowski's inequality, 
	\[\begin{aligned}
		&\lVert\mathcal{L}\mathcal{O}_{G,G_\partial}(\tilde{U})\rVert_{L^2(\Omega)}
		= \norm{\int_{\Omega} \mathcal{L}G(x,\bm{\cdot}) \mathcal{P}(U)(x) \dd{x} + \int_{\Gamma} \mathcal{L}G_{\partial}(x,\bm{\cdot}) \mathcal{B}(U)(x) \dd{S(x)}}_{L^2(\Omega)} \\
		&\qquad\qquad\le \int_{\Omega} \norm{\mathcal{L}G(x,\bm{\cdot})}_{L^2(\Omega)} \abs{\mathcal{P}(U)(x)} \dd{x} + \int_{\Gamma} \norm{\mathcal{L}G_{\partial}(x,\bm{\cdot})}_{L^2(\Omega)} \abs{\mathcal{B}(U)(x)} \dd{S(x)} ,
	\end{aligned}\]
	which is bounded, using the Cauchy--Schwarz inequality, by 
	\[ \norm{\mathcal{P}(U)}_{L^2(\Omega)} \Big\{ \int_{\Omega} \norm{\mathcal{L}G(x,\bm{\cdot})}_{L^2(\Omega)}^{2} \dd{x} \Big\}^{1/2} + \norm{\mathcal{B}(U)}_{L^2(\Gamma)} \Big\{ \int_{\Gamma} \norm{\mathcal{L}G_{\partial}(x,\bm{\cdot})}_{L^2(\Omega)}^{2} \dd{S(x)} \Big\}^{1/2} .\]
	This implies that the tail distribution of $\lVert\mathcal{L}\mathcal{O}_{G,G_\partial}(\tilde{U})\rVert_{L^2(\Omega)}$ is controlled by the tail distributions of $\norm{\mathcal{P}(U)}_{L^2(\Omega)}$ and $\norm{\mathcal{B}(U)}_{L^2(\Gamma)}$ as well as some norm of $\mathcal{O}_{G,G_\partial}$. 
\end{example}

\subsection{Simultaneous diagonalization}
To better express the bias and variance of our estimator, we perform simultaneous diagonalization of the two quadratic forms $\norm{\bm{\cdot}}_{\Sigma}^{2},\norm{\bm{\cdot}}_{\mathcal{H}}^{2}$, allowing us to exploit series expansions of $\mathcal{O}\in\mathcal{H}$. 
We describe it more precisely in the following proposition. 
\begin{proposition}\label{prop:diag}
	Under Assumption~\ref{asm:cov}, there exists a sequence $\varOmega_{1},\varOmega_{2},\dots$ in $\mathcal{H}$ such that 
	$\varSigma\varOmega_{k} = \gamma_{k}\varOmega_{k}$ and $\langle\varOmega_{j},\varOmega_{k}\rangle_{\mathcal{H}} = \gamma_{k}^{-1}\delta_{jk}$ for some $\gamma_{k}\in\mathbb{R}_{+}$, where $\delta_{jk}$ is the Kronecker delta; and that 
	$\mathcal{O} = \sum_{k=1}^{\infty} \gamma_{k}\langle\mathcal{O},\varOmega_{k}\rangle_{\mathcal{H}} \varOmega_{k}$ for any $\mathcal{O}\in\mathcal{H}$, which converges absolutely. 
\end{proposition}
The alignment between $\norm{\bm{\cdot}}_{\Sigma}^{2}$ and $\norm{\bm{\cdot}}_{\mathcal{H}}^{2}$ plays an important role. 
We make the following assumption to quantify the regularity within our model. Note that similar conditions have appeared in \citet{cai2012minimax,stepaniants2023learning}. 
\begin{assumption}\label{asm:eig}
	Let the $\gamma_{k}$ in Proposition~\ref{prop:diag} be in descending order. 
	There exists some constant $r > 1$ such that $\gamma_{k} \asymp k^{-r}$. 
\end{assumption}
Here we show a concrete example to enhance intuition. It is notable that the decay rate $r$ in Assumption~\ref{asm:eig} is jointly determined by $\norm{\bm{\cdot}}_{\Sigma}^{2}$ and $\norm{\bm{\cdot}}_{\mathcal{H}}^{2}$. 
\begin{example}
	Continue from Example~\ref{ex:opRKHS}, and restrict to the case $\mathcal{H}_{\Gamma} = \{0\}$ for simplicity. 
	Suppose that $\mathcal{L}$ admits a spectral decomposition with eigenvalue--eigenfunction pairs $\{(\mu_{j},\psi_{j})\}_{j=1}^{\infty} \subset \mathbb{R}_{+} \times L^2(\Omega)$,
	and that $C_{\mathcal{P}}(x_1,x_2) = \E\{\mathcal{P}(U)(x_1)\mathcal{P}(U)(x_2)\}$ has the eigendecomposition $C_{\mathcal{P}}(x_1,x_2) = \sum_{j=1}^{\infty} \lambda_{j} \varphi_{j}(x_1)\varphi_{j}(x_2)$, where $\{(\lambda_{j},\varphi_{j})\}_{j=1}^{\infty}$ are the eigenpairs. 
	We require $\{\psi_{j}\}_{j=1}^{\infty}$ and $\{\varphi_{j}\}_{j=1}^{\infty}$ to be orthonormal systems of $L^{2}(\Omega)$. 
	For any function $G \in \mathcal{H}_{\Omega}$, write $\mathcal{O}_{G} = \mathcal{O}_{G,0} \in \mathcal{H}$ for brevity, and denote $g_{jj'} = \int_{\Omega^2} G(x,y) \varphi_{j}(x) \psi_{j'}(y) \dd{x}\dd{y}$ so that the expansion $G(x,y) = \sum_{j,j'} g_{jj'} \varphi_{j}(x) \psi_{j'}(y)$ holds. 
	Then for $G^{(1)},G^{(2)} \in \mathcal{H}_{\Omega}$, 
	\[ \langle\mathcal{O}_{G^{(1)}},\mathcal{O}_{G^{(2)}}\rangle_{\Sigma} 
	= \int_{\Omega^3} \mathcal{L}_{y}G^{(1)}(x_1,y)\mathcal{L}_{y}G^{(2)}(x_2,y) C_{\mathcal{P}}(x_1,x_2) \dd{x_1}\dd{x_2}\dd{y} 
	= \sum_{j,j'} \lambda_{j} \mu_{j'}^{2} g_{jj'}^{(1)} g_{jj'}^{(2)} .\]
	If the operator-valued reproducing kernel $\mathcal{K}$ associated with $\mathcal{H}$ is given by 
	\[ \mathcal{K}(\tilde{u},\tilde{v})f = \int_{\Omega^3} K(x,\bm{\cdot},\xi,\eta) \mathcal{P}(v)(x) \mathcal{P}(u)(\xi) f(\eta) \dd{x}\dd{\xi}\dd{\eta} \]
	for some function $K$ expanded as $K(x,y,\xi,\eta) = \sum_{j,j'} \rho_{jj'} \varphi_{j}(x)\psi_{j'}(y) \varphi_{j}(\xi)\psi_{j'}(\eta)$, then 
	\[ \langle\mathcal{O}_{G^{(1)}},\mathcal{O}_{G^{(2)}}\rangle_{\mathcal{H}} = \sum_{j,j'} \rho_{jj'}^{-1} g_{jj'}^{(1)} g_{jj'}^{(2)} .\]
	It follows that $\varSigma$ acts as $\varSigma\mathcal{O}_{G}=\mathcal{O}_{G^\Sigma}$ where $g^{\Sigma}_{jj'} = \lambda_{j} \mu_{j'}^{2} \rho_{jj'} g_{jj'}$, and the pairs $\{(\gamma_{k},\varOmega_{k})\}_{k=1}^{\infty}$ in Proposition~\ref{prop:diag} are $\{(\lambda_{j} \mu_{j'}^{2} \rho_{jj'}, \mathcal{O}_{G_{jj'}})\}_{j,j'=1}^{\infty}$ where $G_{jj'}(x,y) = \lambda_{j}^{-1/2} \mu_{j'}^{-1} \varphi_{j}(x) \psi_{j'}(y)$. 
	In particular, the spectrum of $\varSigma$ is affected by $\rho_{jj'}$. 
	If $K$ is sufficiently smooth, then $\rho_{jj'}$ decays fast enough to yield desired $\gamma_k$. 
	See \citet{yuan2010reproducing,cai2012minimax,zhang2020faster,stepaniants2023learning} for examples of reproducing kernels that have eigenvalues with various rate of decay. 
\end{example}

\subsection{Estimation error and optimality}
Now we investigate the asymptotic properties of the regularization estimator \eqref{eq:est}. 

To begin with, define the oracle $\bar{\mathcal{T}}_{\lambda} = \operatorname{arg\,min}_{\mathcal{O}\in\mathcal{H}} P\ell_{\mathcal{O};\lambda}$. 
Under Assumption~\ref{asm:cov}, it is immediate that $\bar{\mathcal{T}}_{\lambda} = (\varSigma+\lambda\mathcal{I})^{-1}\varSigma\mathcal{T}$, where $\mathcal{I}$ is the identity operator on $\mathcal{H}$. 
The difference between $\hat{\mathcal{T}}_{\lambda}$ and $\bar{\mathcal{T}}_{\lambda}$ will be approximated by a Newton--Raphson update. 
For this purpose, the first and second order Fr\'{e}chet derivatives of the regularized loss \eqref{eq:loss} are calculated as 
\[ \dot{\ell}_{\mathcal{O};\lambda}[\mathcal{O}_{1}](\tilde{u},f) = - 2 \langle f-\mathcal{L}\mathcal{O}(\tilde{u}) , \mathcal{L}\mathcal{O}_{1}(\tilde{u}) \rangle_{L^2(\Omega)} + 2 \lambda \langle\mathcal{O},\mathcal{O}_{1}\rangle_{\mathcal{H}} ,\]
\[ \ddot{\ell}_{\mathcal{O};\lambda}[\mathcal{O}_{1},\mathcal{O}_{2}](\tilde{u},f) = 2 \langle \mathcal{L}\mathcal{O}_{1}(\tilde{u}) , \mathcal{L}\mathcal{O}_{2}(\tilde{u}) \rangle_{L^2(\Omega)} + 2 \lambda \langle\mathcal{O}_{1},\mathcal{O}_{2}\rangle_{\mathcal{H}} .\]
By the Riesz representation theorem, $\dot{\ell}_{\mathcal{O};\lambda}(\tilde{u},f) : \mathcal{O}_{1} \mapsto \dot{\ell}_{\mathcal{O};\lambda}[\mathcal{O}_{1}](\tilde{u},f)$ can be viewed as an element in $\mathcal{H}$, and $\ddot{\ell}_{\mathcal{O};\lambda}(\tilde{u},f)$ an operator on $\mathcal{H}$. 
Note that $P\ddot{\ell}_{\mathcal{O};\lambda} = 2(\varSigma+\lambda\mathcal{I})$ for any $\mathcal{O}$. 
Since $\bar{\mathcal{T}}_{\lambda}$ is a minimizer, we have $P\dot{\ell}_{\bar{\mathcal{T}}_{\lambda};\lambda} = 0$, and thus with $\dot{\ell}_{\bar{\mathcal{T}}_{\lambda}} = \dot{\ell}_{\bar{\mathcal{T}}_{\lambda};0}$, 
\[ P_{n}\dot{\ell}_{\bar{\mathcal{T}}_{\lambda};\lambda} = P_{n}\dot{\ell}_{\bar{\mathcal{T}}_{\lambda};\lambda} - P\dot{\ell}_{\bar{\mathcal{T}}_{\lambda};\lambda} = P_{n}\dot{\ell}_{\bar{\mathcal{T}}_{\lambda}} - P\dot{\ell}_{\bar{\mathcal{T}}_{\lambda}} = (P_{n}-P)\dot{\ell}_{\bar{\mathcal{T}}_{\lambda}} .\]

Then we establish a Bahadur representation for the proposed estimator, which is useful for subsequent analysis. Let 
\[ \Delta_{n\lambda} = (P\ddot{\ell}_{\bar{\mathcal{T}}_{\lambda};\lambda})^{-1} P_{n}\dot{\ell}_{\bar{\mathcal{T}}_{\lambda};\lambda} = \{2(\varSigma+\lambda\mathcal{I})\}^{-1}(P_{n}-P)\dot{\ell}_{\bar{\mathcal{T}}_{\lambda}} .\]
\begin{theorem}\label{thm:bahadur}
	Suppose that Assumptions~\ref{asm:err}--\ref{asm:eig} hold. 
	Assume that $\lambda \to 0$ and $n^{-1}\lambda^{-c-1/r} \to 0$ for some constant $1/r < c \le 1$. 
	Then the estimator $\hat{\mathcal{T}}_{\lambda}$ in \eqref{eq:est} satisfies that 
	\[ \lVert \hat{\mathcal{T}}_{\lambda} - \bar{\mathcal{T}}_{\lambda} + \Delta_{n\lambda} \rVert_{\Sigma} = O_{\Pr}(n^{-1}\lambda^{-c/2-1/r}) .\]
\end{theorem}

As a prominent implication of Theorem~\ref{thm:bahadur}, we show an upper bound for the convergence rate of our proposed estimator. 
The convergence rate is in fact optimal, for which we also provide a minimax lower bound. To this end, denote by $\mathcal{F}$ the collection of the distribution $P$ of $(U,F)$ satisfying Assumptions~\ref{asm:err}--\ref{asm:eig} and that $\norm{\mathcal{T}}_{\mathcal{H}} \le C_{\mathcal{T}}$ for some constant $C_{\mathcal{T}} > 1$. 
The following result gives the optimal convergence rate for the regularized estimator $\hat{\mathcal{T}}_{\lambda}$ with an appropriately chosen tuning parameter $\lambda$. 
\begin{theorem}\label{thm:rate}
	Under Assumptions~\ref{asm:err}--\ref{asm:eig}, by setting $\lambda \asymp n^{-r/(r+1)}$, the estimator $\hat{\mathcal{T}}_{\lambda}$ in \eqref{eq:est} satisfies that 
	\[ \lVert \hat{\mathcal{T}}_{\lambda} - \mathcal{T} \rVert_{\Sigma}^{2} = O_{\Pr}\{n^{-r/(r+1)}\} .\]
	There exists a constant $c_{0} > 0$ such that 
	\[ \liminf_{n\to\infty} \inf_{\hat{\mathcal{T}}} \sup_{P\in\mathcal{F}} \Pr_{P}\{ \lVert\hat{\mathcal{T}}-\mathcal{T}\rVert_{\Sigma}^{2} > c_{0} n^{-r/(r+1)} \} > 0 ,\]
	where $\inf_{\hat{\mathcal{T}}}$ is taken over all possible estimators $\hat{\mathcal{T}}$ based on the observations $(U_{i},F_{i})$, $i=1,\dots,n$, generated from the model \eqref{eq:model_trans}. 
\end{theorem}
In Theorem~\ref{thm:rate}, the measure of estimation accuracy is related to the excess prediction risk. To see this, for any $\mathcal{O}\in\mathcal{H}$, we have 
\[\begin{aligned}
	\norm{\mathcal{O}-\mathcal{T}}_{\Sigma}^{2} 
	&= \E\{\lVert \mathcal{L}(\mathcal{O}-\mathcal{T})(\tilde{U}) \rVert_{L^{2}(\Omega)}^{2}\} \\
	&= \E\{\lVert \varepsilon - \mathcal{L}(\mathcal{O}-\mathcal{T})(\tilde{U}) \rVert_{L^{2}(\Omega)}^{2}\} - \E\{\lVert\varepsilon\rVert_{L^{2}(\Omega)}^{2}\} \\
	&= \E\{\lVert F - \mathcal{L}\mathcal{O}(\tilde{U}) \rVert_{L^{2}(\Omega)}^{2}\} - \E\{\lVert F - \mathcal{L}\mathcal{T}(\tilde{U}) \rVert_{L^{2}(\Omega)}^{2}\} .
\end{aligned}\]
Therefore, the difficulty of the prediction problem is characterized by the convergence rate $n^{-r/(r+1)}$ as the sample size $n$ increases. As noticed in Assumption~\ref{asm:eig}, the decay rate $r$ of $\gamma_{k}$'s reflects the combined effect of $\norm{\bm{\cdot}}_{\Sigma}^{2}$ and $\norm{\bm{\cdot}}_{\mathcal{H}}^{2}$, which brings the price we need to pay for performing valid statistical inference.

\subsection{Asymptotics of goodness-of-fit test}
Next, we analyze the asymptotic properties of the proposed goodness-of-fit test statistic $Q_n$ defined in \eqref{eq:test}, which further leads to the consistency. 
For the parametric model in $H_0$, we make the following assumptions. 
\begin{assumption}\label{asm:para-Lip}
	The parametric family of differential operators $\{\mathcal{D}_{\theta}:\theta\in\Theta\}$ satisfies that $\norm{\mathcal{D}_{\theta_1}(u)-\mathcal{D}_{\theta_2}(u)}_{L^{2}(\Omega)} \le \dot{\mathcal{D}}(u) \norm{\theta_{1}-\theta_{2}}$ for any $\theta_{1},\theta_{2}\in\Theta$ and $u \in L^{2}(\Omega)$, where $\dot{\mathcal{D}}$ is a measurable operator on $L^{2}(\Omega)$ such that $\E\{\lVert\dot{\mathcal{D}}(U)\rVert_{L^2(\Omega)}^{2}\} < \infty$. 
\end{assumption}
\begin{assumption}\label{asm:para-size}
	On an event $E_n$ with $\Pr(E_n) \to 1$ as $n \to \infty$, the estimator $\hat{\theta}$ lies in a compact subset $\Theta_{n}$ of $\Theta$ such that $\theta_{0}\in\Theta_{n}$ and $\operatorname{diam}(\Theta_{n}) \lesssim n^{-1/2}$, where $\operatorname{diam}(\Theta_{n}) = \sup_{\theta_{1},\theta_{2}\in\Theta_{n}} \norm{\theta_{1}-\theta_{2}}$ is the diameter of $\Theta_n$. 
\end{assumption}
Assumption~\ref{asm:para-Lip} is commonly used in parametric inference, measuring the continuity of the statistical model with respect to the parameter of interest. Assumption~\ref{asm:para-size} characterizes the accuracy of the null estimator $\hat{\theta}$, using the classical parametric convergence rate $n^{-1/2}$. Such a setup provides a standard workplace for parameter statistics, and hence the estimation error on the null model is negligible when we derive results for nonparametric testing. 

To understand the testing boundary, we establish the asymptotic normality for the test statistic in the case of an appropriate deviation from the null hypothesis. 
For convenience of notation, define $\mathcal{T}_{0} = \mathcal{L}^{-1}\circ\mathcal{D}_{\theta_0}\circ\mathcal{A}$ as \eqref{eq:operator}, and let \[ \mathcal{Q} = \mathcal{T} - \mathcal{T}_{0} = \mathcal{L}^{-1}\circ(\mathcal{D}-\mathcal{D}_{\theta_0})\circ\mathcal{A} .\] 
\begin{theorem}\label{thm:test}
	Assume that Assumptions~\ref{asm:err}--\ref{asm:para-size} hold, $\lambda \to 0$ and $n^{-1}\lambda^{-c-1/r} \to 0$ for some constant $1/r < c \le 1$. Then the statistic $Q_n$ defined in \eqref{eq:test} satisfies that 
	\[ (Q_n - \lVert(\varSigma+\lambda\mathcal{I})^{-1}\varSigma\mathcal{Q}\rVert_{\Sigma}^{2} - \mu_{n}) / \sigma_{n} \xrightarrow{d} \mathcal{N}(0,1) ,\]
	provided $\mathcal{Q} \in \mathcal{H}$ with $\lVert(\varSigma+\lambda\mathcal{I})^{-1}\varSigma\mathcal{Q}\rVert_{\Sigma}^{2} \lesssim n^{-1}\lambda^{-1/r}$, where $\mu_{n} \asymp \sigma_{n} \asymp n^{-1}\lambda^{-1/r}$ not depending on $\mathcal{Q}$. 
\end{theorem}
Theorem~\ref{thm:test} shows that the power of the goodness-of-fit test based on $Q_n$ is asymptotically constant on regions of $\mathcal{D}$ with a fixed $\lVert(\varSigma+\lambda\mathcal{I})^{-1}\varSigma\mathcal{Q}\rVert_{\Sigma}^{2}$, and the scale of such regions are of order $n^{-1}\lambda^{-1/r}$. The consistency of the test is achieved when $n\lambda^{1/r} \to \infty$, which is often satisfied for a suitable tuning parameter $\lambda$ as the sample size $n$ grows, e.g., the optimal choice $\lambda \asymp n^{-r/(r+1)}$ in Theorem~\ref{thm:rate}. 

Since the normal approximation in Theorem~\ref{thm:test} involves some quantities $\mu_{n},\sigma_{n}$ that are difficult to evaluate, we appeal to a bootstrap procedure defined in \eqref{eq:btsp} to obtain critical values. Then we show that the bootstrapped version of the test statistic has similar asymptotic behavior. 
Denote by $\mathscr{L}^{*}(S)$ the conditional distribution of a statistic $S$ given the observations $(U_{i},F_{i})$, $i=1,\dots,n$. 
Let $d$ be any distance on the space of probability measures on $\mathbb{R}$, e.g., the total variation distance $d(\pi_{1},\pi_{2}) = \sup_{A\subset\mathbb{R}}\abs{\pi_{1}(A)-\pi_{2}(A)}$. 
\begin{theorem}\label{thm:btsp}
	Assume that Assumptions~\ref{asm:err}--\ref{asm:para-size} hold, $\lambda \to 0$ and $n^{-1}\lambda^{-c-1/r} \to 0$ for some constant $1/r < c \le 1$. Then the statistic $Q_n^b$ defined in \eqref{eq:btsp} satisfies that 
	\[ d[\mathscr{L}^{*}\{(Q_{n}^{b}-\mu_{n})/\sigma_{n}\}, \mathcal{N}(0,1)] \xrightarrow{\Pr} 0,\]
	provided $\mathcal{Q} \in \mathcal{H}$ with $\lVert(\varSigma+\lambda\mathcal{I})^{-1}\varSigma\mathcal{Q}\rVert_{\Sigma}^{2} \lesssim n^{-1}\lambda^{-1/r}$, where $\mu_{n}, \sigma_{n}$ are given in Theorem~\ref{thm:test}. 
\end{theorem}
Theorem~\ref{thm:btsp} implies the validity of the bootstrap, since quantiles of $\mathscr{L}^{*}(Q_{n}^{b})$ serves as good approximations of quantiles of the null distribution of $Q_{n}$.

\section{Simulation Studies}
\label{sec:simulation}
In this section, we explore the numerical performance of the proposed procedures for inference for the function-on-function differential model \eqref{eq:model}. 
We simulate data with the following settings. 
The domain $\Omega$ is the one-dimensional interval $[0,1]$, whose boundary is $\Gamma = \{0,1\}$. 
The predictor functions are $U_{i} = \sum_{k=1}^{10} k^{-3} Z_{ik} \phi_{k}$, $i=1,\dots,n$, where $\phi_{k}(x) = \sqrt{2} \cos(k\pi x)$, and $Z_{ik}$'s are independent random variables uniformly distributed on $(-\sqrt{3},\sqrt{3})$. In particular, each $Z_{ik}$ has zero mean and unit variance.
The differential operator is given by $\mathcal{D} = - \nabla^{2} - \omega^{2}$ that arises in the one-dimensional Helmholtz equation, where $\nabla^{2}$ is the Laplacian operator and $\omega \ge 0$ denotes the wavenumber. 
Note that it is for computational convenience that $\phi_k$'s are taken to be eigenfunctions of $\mathcal{D}$, i.e., $\mathcal{D}\phi_{k} = (k\pi)^{2} + \omega^{2}$. 
The observational errors are $\varepsilon_{i} = \sigma \sum_{k=1}^{10} \epsilon_{ik} \phi_{k}$, $i=1,\dots,n$, where $\epsilon_{ik}$'s are independent random variables generated from the standard normal distribution $\mathcal{N}(0,1)$, and $\sigma > 0$ is chosen to control the signal-to-noise ratio defined as $\mathrm{SNR} = \big[\E\{\norm{\mathcal{D}(U_{i})}_{L^{2}(\Omega)}^{2}\}/\E\{\norm{\varepsilon_{i}}_{L^{2}(\Omega)}^{2}\}\big]^{1/2}$. 
We generate samples with $n = 200,400$, $\mathrm{SNR} = 8,3,1$, and various $\omega$. 

The transformation from \eqref{eq:model} to \eqref{eq:model_trans} is realized using the differential operators $\mathcal{L} = \mathcal{P} = -\nabla^{2}$ on $\Omega$ and the Dirichlet boundary condition $\mathcal{B} = \mathcal{I}$ on $\Gamma$. 
To proceed with the operator reproducing kernel Hilbert space framework, we focus on Example~\ref{ex:opRKHS} with $K(x,y,\xi,\eta) = K_{1}(x,\xi)K_{1}(y,\eta)$ and $K_{\partial}(x,y,\xi,\eta) = \delta(x-\xi)K_{1}(y,\eta)$, where $K_{1}(x,y) = (2\pi h^{2})^{-1/2} \exp\{-(x-y)^{2}/(2h^{2})\}$ is the Gaussian kernel with bandwidth $h = 0.01$, and $\delta(\cdot)$ is the Dirac delta function. 
For comparison and model checking, we also introduce a parametric family of differential operators, namely, $\mathcal{D}_{\theta} = - \theta \nabla^{2}$, $\theta \in \mathbb{R}_{+}$. The estimate $\hat{\theta}$ for $\theta$ is obtained using least squares, i.e., $\hat{\theta} = \operatorname{arg\,min}_{\theta\in\mathbb{R}_{+}}\sum_{i=1}^{n}\norm{F_{i}-\mathcal{D}_{\theta}(U_{i})}_{L^2(\Omega)}^{2}$. 

To assess the estimation accuracy, we consider the error sum of squares defined as $\sum_{i=1}^{n}\lVert(\hat{\mathcal{D}}-\mathcal{D})(U_{i})\rVert_{L^{2}(\Omega)}^{2}$ for any estimator $\hat{\mathcal{D}}$ for the underlying differential operator $\mathcal{D}$, and we denote it by $\mathrm{ESS}_{\lambda}$ and $\mathrm{ESS}_{\hat{\theta}}$ for the proposed regularization estimator $\hat{\mathcal{D}}_{\lambda}$ and the parametric estimator $\mathcal{D}_{\hat{\theta}}$, respectively. 
The tuning parameter $\lambda$ is selected by minimizing the generalized cross-validation criterion, \[ \mathrm{GCV}_{\lambda} = n^{-1} \mathrm{RSS}_{\lambda} / \{1 - n^{-1}\tr(\mathcal{S}_{\lambda})\}^{2} ,\] where $\mathrm{RSS}_{\lambda}$ is the residual sum of squares given by $\mathrm{RSS}_{\lambda} = \sum_{i=1}^{n}\lVert F_{i} - \hat{\mathcal{D}}_{\lambda}(U_{i})\rVert_{L^{2}(\Omega)}^{2}$, and $\tr(\mathcal{S}_{\lambda})$ is the trace of the smoothing operator $\mathcal{S}_{\lambda}$. Using the matrix representation $\mathbf{S}_{\lambda}$ in \eqref{eq:smooth}, one has $\tr(\mathcal{S}_{\lambda}) = \tr(\mathbf{S}_{\lambda}) / p$. 
The results under different choices of $\lambda$ are presented in the following Table~\ref{tab:tuning}, where for space economy we fix $n = 200$ and $\mathrm{SNR} = 3$. 
It can be seen that as $\lambda$ increases, the error first increases and then decreases, reflecting the trade-off between bias and variance, and that the selection procedure is valid. 

\begin{table}[!ht]
	\caption{ESS and GCV with varying tuning parameters. Reported are the average and standard deviation (in the parenthesis) based on 1000 Monte Carlo replications in the case where $n = 200$ and $\mathrm{SNR} = 3$.}
	\label{tab:tuning}
	\centering
	\begin{tabular}{|c|crrrrrr|}
		\hline
		$\omega$ & $\lambda$ & $10^0$ & $10^1$ & $10^2$ & $10^3$ & $10^4$ & $10^5$ \\
		\hline \multirow{4}{*}{0} 
		& $\mathrm{ESS}_{\lambda}$ & 167.9~ & 167.8~ & 166.7~ & \textbf{162.8} & 307.1~ & 3715~ \\
		& & (0.7) & (0.7) & (0.7) & (0.7) & (1.1) & (3.3) \\
		& $\mathrm{GCV}_{\lambda}$ & 17.68~ & 17.68~ & 17.67~ & \textbf{17.65} & 18.43~ & 36.68~ \\
		& &  (0.02) & (0.02) & (0.02) &  (0.02) &  (0.02) & (0.02) \\
		\hline \multirow{4}{*}{0.84} 
		& $\mathrm{ESS}_{\lambda}$ & 185.2~ & 185.1~ & 183.9~ & \textbf{179.4} & 331.0~ & 4084~ \\
		& & (0.8) & (0.8) & (0.8) & (0.8) &  (1.2) & (3.7) \\
		& $\mathrm{GCV}_{\lambda}$ & 19.50~ & 19.50~ & 19.49~ & \textbf{19.47} & 20.29~ & 40.39~ \\
		& & (0.02) & (0.02) & (0.02) & (0.02) &  (0.02) & (0.03)  \\
		\hline
	\end{tabular}
\end{table}

We then turn to the comparison of the proposed estimator $\hat{\mathcal{D}}_{\lambda}$ and the parametric estimator $\mathcal{D}_{\hat{\theta}}$. 
The results are summarized in Table~\ref{tab:compare}, where we use various sample sizes $n$, signal-to-noise ratios $\mathrm{SNR}$, and  wavenumbers $\omega$. 
In every scenario, the total sum of squares, defined as $\mathrm{TSS} = \sum_{i=1}^{n} \norm{\mathcal{D}(U_{i})}_{L^{2}(\Omega)}^{2}$, is of order $10^4$ to $10^5$. 
For both estimators, the $\mathrm{ESS}$ turns out to be a small portion of $\mathrm{TSS}$, implying that the model fits the data well. 
We first focus on the case $\omega = 0$, whence the true differential operator lies in the parametric model. 
As expected, the $\mathrm{ESS}$ is roughly proportional to $\mathrm{SNR}^{-2}$ when $n$ is fixed, which corresponds to the magnitude of the variance of the observational noise. 
If $\mathrm{SNR}$ is given while $n$ is adjusted from $200$ to $400$, then $\mathrm{TSS}$ becomes approximately doubled, but notably, the $\mathrm{ESS}$ of both estimators remain the same scale. This is accounted for by the consistency of the estimators. It is worth mentioning that here the nonparametric convergence rate is close to the parametric one, partly because the Gaussian kernel we adopted is sufficiently smooth, which does not violate Theorem~\ref{thm:rate}. 
Then we let $\omega$ vary while keeping $n$ and $\mathrm{SNR}$ fixed. The larger $\omega$ implies more significant deviation from the parametric model, and thus $\mathrm{ESS}_{\hat{\theta}}$ inflates, suggesting that the specific form of differential operators $\mathcal{D}_{\theta}$ is not suitable. 
On the other hand, $\mathrm{ESS}_{\lambda}$ exhibits relatively stable behavior, since the proposed estimator $\hat{\mathcal{D}}_{\lambda}$ enjoys versatility in dealing with a wide range of situations. 
These results underscore the adaptability and efficacy of the proposed method of estimation. 

\begin{table}[!ht]
	\caption{Comparison of the proposed estimator and the parametric estimator. The proposed estimates were computed with tuning parameter $\lambda$ chosen to minimize the $\mathrm{ESS}_{\lambda}$. Reported are the average and standard deviation (in the parenthesis) based on 1000 Monte Carlo replications.}
	\label{tab:compare}
	\centering
	\begin{tabular}{|c|crrrr|}
		\hline
		$n = 200$ & $\omega$ & 0 & 0.50 & 1.00 & 1.50 \\
		\multirow{2}{*}{$\mathrm{SNR} = 8$} & $\mathrm{ESS}_{\lambda}$ & 23.4(0.1) & 24.3(0.1) & 26.9(0.1) & 31.7(0.1) \\ 
		& $\mathrm{ESS}_{\hat{\theta}}$ & 0.2(0.01) & 3.5(0.01) & 52.5(0.1) & 264.6(0.2) \\
		\hline
		$n = 200$ & $\omega$ & 0 & 0.84 & 1.68 & 2.52 \\
		\multirow{2}{*}{$\mathrm{SNR} = 3$} & $\mathrm{ESS}_{\lambda}$ & 162.8(0.7) & 179.4(0.8) & 235.6(1.0) & 350.8(1.5) \\ 
		& $\mathrm{ESS}_{\hat{\theta}}$ & 1.7(0.1) & 27.9(0.1) & 418.4(0.4) & 2109(1.8) \\
		\hline
		$n = 200$ & $\omega$ & 0 & 1.66 & 3.32 & 4.98 \\
		\multirow{2}{*}{$\mathrm{SNR} = 1$} & $\mathrm{ESS}_{\lambda}$ & 1374(6) & 1966(9) & 4564(21) & 11632(53) \\ 
		& $\mathrm{ESS}_{\hat{\theta}}$ & 16(1) & 419(1) & 6395(6) & 32244(28) \\
		\hline
		$n = 400$ & $\omega$ & 0 & 0.42 & 0.84 & 1.26 \\
		\multirow{2}{*}{$\mathrm{SNR} = 8$} & $\mathrm{ESS}_{\lambda}$ & 23.3(0.1) & 23.9(0.1) & 25.7(0.1) & 28.9(0.1) \\ 
		& $\mathrm{ESS}_{\hat{\theta}}$ & 0.2(0.01) & 3.5(0.01) & 52.3(0.03) & 263.8(0.2) \\
		\hline
		$n = 400$ & $\omega$ & 0 & 0.70 & 1.40 & 2.10 \\
		\multirow{2}{*}{$\mathrm{SNR} = 3$} & $\mathrm{ESS}_{\lambda}$ & 163.2(0.7) & 174.7(0.8) & 212.3(0.9) & 284.9(1.2) \\ 
		& $\mathrm{ESS}_{\hat{\theta}}$ & 1.5(0.1) & 26.7(0.1) & 403.6(0.3) & 2036(1.3) \\
		\hline
		$n = 400$ & $\omega$ & 0 & 1.32 & 2.64 & 3.96 \\
		\multirow{2}{*}{$\mathrm{SNR} = 1$} & $\mathrm{ESS}_{\lambda}$ & 1395(6) & 1759(8) & 3187(14) & 6655(30) \\ 
		& $\mathrm{ESS}_{\hat{\theta}}$ & 13(1) & 334(1) & 5110(4) & 25778(16) \\
		\hline
	\end{tabular}
\end{table}

Next, we consider the goodness-of-fit test for the parametric model $\{\mathcal{D}_{\theta}\}$. 
Let the nominal significance level be $\alpha = 5\%$. 
To conduct the proposed test for each sample, the bootstrap procedure \eqref{eq:btsp} is performed $B=1000$ times. 
Since the vector of parametric residuals $(\tilde{\varepsilon}_{i})$ and the vector of nonparametric residuals $(\hat{\varepsilon}_{i})$ are both available, we could use either one for bootstrap calculation. Besides, we introduce a mixed strategy where $(\tilde{\varepsilon}_{i})$ weights $1/3$ and $(\hat{\varepsilon}_{i})$ weights $2/3$. 
In Table~\ref{tab:test_tuning}, we record the rejection proportions for the three bootstrap methods based on 1000 Monte Carlo replications in a null case. 
When the tuning parameter $\lambda$ is near or less than the optimal choice for estimation, the rejection proportions present rather small changes within any bootstrap method. 
The bootstrap using only parametric (resp.\ nonparametric) residuals turns out to be conservative (resp.\ radical), while the mixed strategy yields satisfactory performance where the empirical size is around the nominal level. 
Therefore, we recommend performing bootstrap with mixed residuals, and we adopt this manner in what follows. 

\begin{table}[!ht]
	\caption{Rejection proportions (\%) calculated for different tuning parameters based on 1000 Monte Carlo replications in the case where $n = 200$, $\mathrm{SNR} = 3$, and $\omega = 0$. The bootstrap procedure performed using parametric, nonparametric, and mixed residuals is indicated by para., nonp., and mix., respectively.}
	\label{tab:test_tuning}
	\centering
	\begin{tabular}{cccccccccc}
		\hline
		$\lambda$ & $10^1$ & $10^2$ & $10^{2.6}$ & $10^{2.8}$ & $10^3$ & $10^{3.2}$ & $10^{3.4}$ & $10^4$ & $10^5$ \\
		para. &  1.7 & 1.9 & 2.0 & 2.2 & 2.1 & 2.6 & 2.6 & 2.8 & 3.3 \\ 
		nonp. &  7.1 & 6.8 & 6.8 & 7.0 & 6.9 & 7.2 & 6.6 & 6.1 & 0.0 \\ 
		mix. &  5.1 & 4.7 & 5.1 & 4.8 & 5.1 & 5.3 & 5.0 & 4.3 & 0.4 \\
		\hline
	\end{tabular}
\end{table}

\begin{figure}[!ht]
	\centering
	\includegraphics[width=0.76\linewidth]{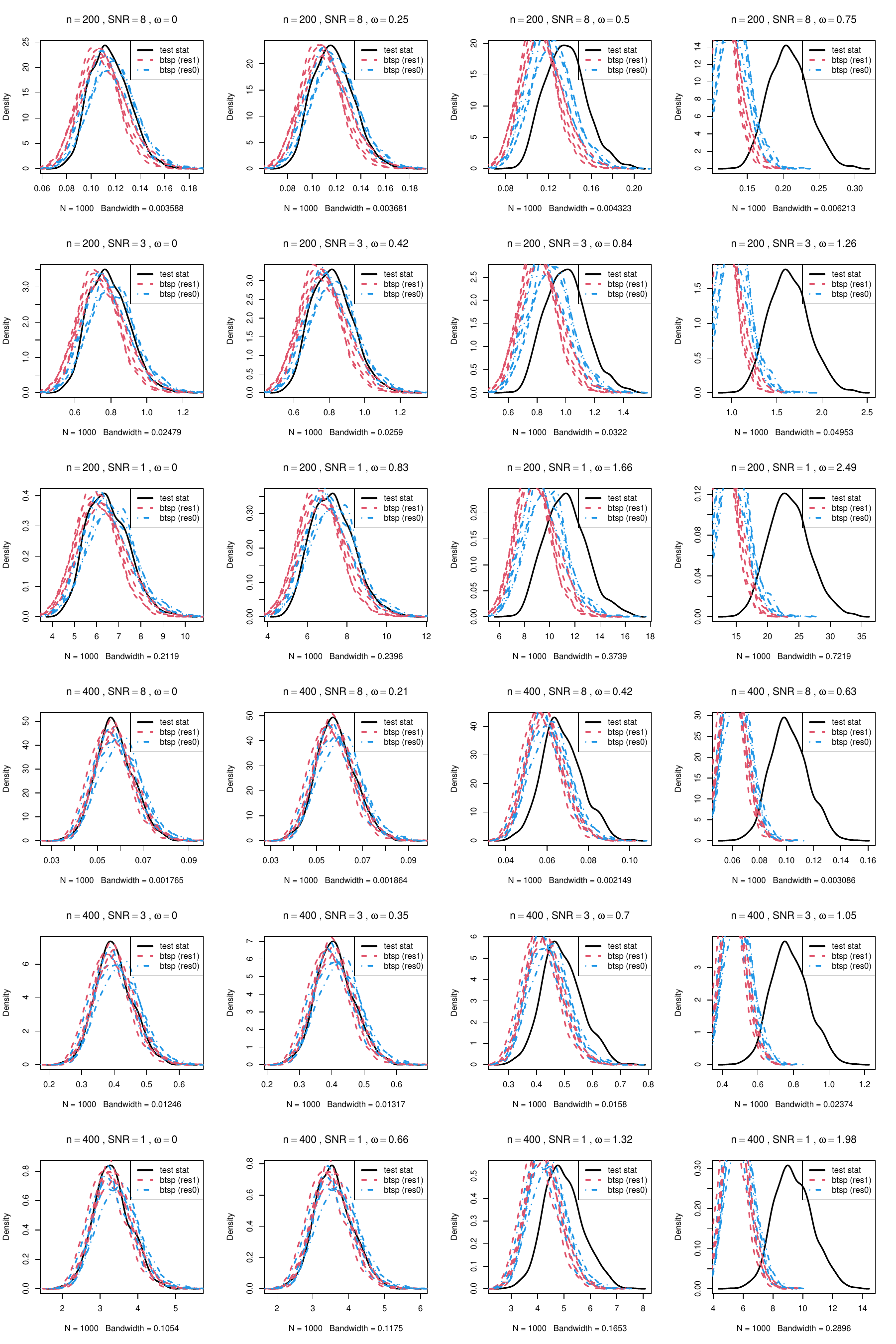}
	\caption{Monte Carlo densities of the test statistic (solid line) and its bootstrap for several samples (dashed line). The red and blue lines correspond to bootstrap using nonparametric and parametric residuals, respectively.}
	\label{fig:stat}
\end{figure}

Regarding power performance under alternatives, we let $\omega$ increase to deliver deviation from the null hypothesis, and calculate the rejection proportions under various settings of $n$ and $\mathrm{SNR}$. 
The results are shown in Table~\ref{tab:power}. 
We see that the empirical power is achieved for large enough $\omega$, demonstrating the consistency of the proposed test. 
Small sample sizes and small signal-to-noise ratios could hinder the detection of lack of fit, in which case a larger $\omega$ is required to attain a given rejection proportion. 
For illustration, we plot the estimated densities of the test statistic and its bootstrap in Figure~\ref{fig:stat}. The asymptotic normality properties claimed in Theorems \ref{thm:test} and \ref{thm:btsp} are numerically supported. 
When $\omega$ is small, the finite sample distribution of the test statistic is close to the bootstrap version. As $\omega$ increases, the test statistic becomes more affected by its bias, thus tending to reject the null hypothesis. 

\begin{table}[!ht]
	\caption{Rejection proportions (\%) calculated in different scenarios based on 1000 Monte Carlo replications. The test statistics were computed with tuning parameter $\lambda$ chosen to minimize the $\mathrm{ESS}_{\lambda}$.}
	\label{tab:power}
	\centering
	\begin{tabular}{|c|crrrrr|}
		\hline
		$n = 200$ & $\omega$ & 0 & 0.25 & 0.50 & 0.75 & 1.00 \\
		$\mathrm{SNR} = 8$ & rej. & 5.1 & 5.7 & 22.7 & 97.7 & 100.0 \\
		\hline
		$n = 200$ & $\omega$ & 0 & 0.42 & 0.84 & 1.26 & 1.68 \\
		$\mathrm{SNR} = 3$ & rej. & 5.1 & 5.9 & 24.7 & 97.3 & 100.0 \\
		\hline
		$n = 200$ & $\omega$ & 0 & 0.83 & 1.66 & 2.49 & 3.32 \\
		$\mathrm{SNR} = 1$ & rej. & 5.1 & 6.4 & 35.9 & 97.4 & 100.0 \\
		\hline
		$n = 400$ & $\omega$ & 0 & 0.21 & 0.42 & 0.63 & 0.84 \\
		$\mathrm{SNR} = 8$ & rej. & 4.6 & 5.5 & 22.6 & 97.9 & 100.0 \\
		\hline
		$n = 400$ & $\omega$ & 0 & 0.35 & 0.70 & 1.05 & 1.40 \\
		$\mathrm{SNR} = 3$ & rej. & 4.4 & 5.4 & 23.7 & 97.9 & 100.0 \\
		\hline
		$n = 400$ & $\omega$ & 0 & 0.66 & 1.32 & 1.98 & 2.64 \\
		$\mathrm{SNR} = 1$ & rej. & 4.2 & 5.6 & 28.5 & 98.3 & 100.0 \\
		\hline
	\end{tabular}
\end{table}

\section{Real Data Example}
\label{sec:realdata}
Here we apply the proposed method to an ERA5 dataset studied by \citet{rothlisberger2023quantifying}, available at {\sf https://doi.org/10.3929/ethz-b-000571107}. The ERA5 is the latest reanalysis data of the European Centre for Medium-Range Weather Forecasts. We consider the TX1day trajectories started at $10\,\mathrm{hPa}$ above ground within November 2020. Each trajectory is stored with a 3-hourly temporal resolution during 15 days, resulting in 121 measurement points which we denote by $t$. 
For each $t$, the following variables are traced: pressure $p$ in $\mathrm{hPa}$, temperature $T_\mathrm{real}$ in $\mathrm{K}$, and potential temperature $T_\mathrm{pot}$ in $\mathrm{K}$. 
These variables are known to satisfy the thermodynamic energy equation \citep[equation (3)]{rothlisberger2023quantifying}
\begin{equation*}
	\dv{T_{\mathrm{real}}}{t} = \frac{\kappa T_{\mathrm{real}}}{p}\dv{p}{t} + \Big(\frac{p}{p_0}\Big)^{\kappa} \dv{T_{\mathrm{pot}}}{t} ,
\end{equation*}
where $\kappa = 0.286$ is the Poisson constant for gas, and $p_{0} = 1000 \,\mathrm{hPa}$ is the reference pressure. 
For simplicity, we rewrite the above equation as 
\begin{equation}\label{eq:thermo}
	\Big(\frac{p}{p_0}\Big)^{-\kappa} \Big(\dv{T_{\mathrm{real}}}{\log(p)} - \kappa T_{\mathrm{real}}\Big) = \dv{T_{\mathrm{pot}}}{\log(p)} .
\end{equation}
The left hand-side of \eqref{eq:thermo} is thought of as the response function, and $T_{\mathrm{pot}}$ the predictor function. 
We are interested in the verification of the differential operator $\mathcal{D}_{1} = \dv{\log(p)}$. 

\begin{figure}[!ht]
	\centering
	\includegraphics[width=0.8\linewidth]{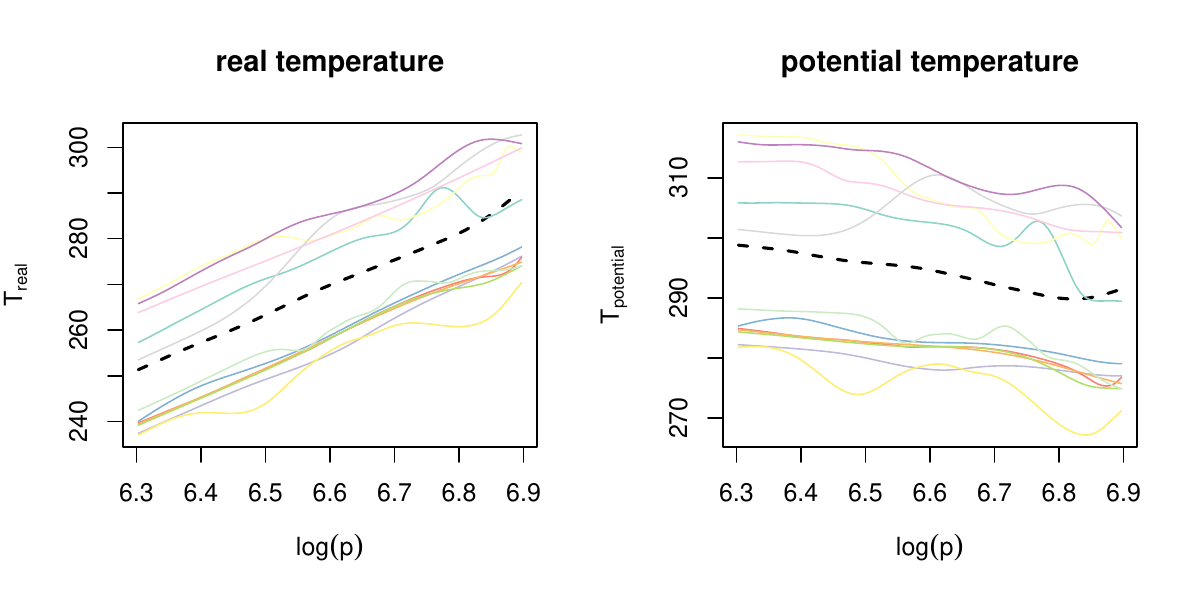}
	\caption{Temperatures versus logarithm of pressure in an ERA5 dataset. Each solid line corresponds to a subject, while the dashed line represents the sample mean.}
	\label{fig:TX1}
\end{figure}

To begin with, the data are preprocessed as follows. 
We choose the trajectories such that the smallest pressure lies in $(\mathrm{e}^{6.29},\mathrm{e}^{6.31})$ and the largest pressure lies in $(\mathrm{e}^{6.89},\mathrm{e}^{6.91})$, and use spline smoothing to express $T_{\mathrm{real}}$ and $T_{\mathrm{pot}}$ as functions of $\log(p) \in (6.3, 6.9)$. 
After excluding the subjects with too large $\abs{\dv{T_{\mathrm{pot}}}{\log(p)}}$, we obtain a sample of size $n = 238$. 
The resulted functions $T_{\mathrm{real}}$ and $T_{\mathrm{pot}}$ on $\Omega = (6.3, 6.9)$ are depicted in Figure~\ref{fig:TX1}. 

We then calculate the response and predictor functions, $(F_{i},U_{i}),\ i=1,\dots,n$, indicated by \eqref{eq:thermo}, where empirical centralization is applied to help us focus on the variation. The centralization step does not affect the dependence structure between $F_i$ and $U_i$, since the operator $\mathcal{D}_{1}$ is supposed to be linear. 
We have $\sum_{i=1}^{n}\norm{F_{i}}_{L^{2}(\Omega)}^{2} = 432680.97$ and $\mathrm{RSS}_{0} = \sum_{i=1}^{n}\norm{F_{i}-\mathcal{D}_{1}(U_{i})}_{L^{2}(\Omega)}^{2} = 296.81$, signaling that the null fit is quite good. 

To conduct the proposed method of estimation and goodness-of-fit testing, we invoke the differential operators $\mathcal{L} = \mathcal{P} = \mathcal{D}_1$ on $\Omega = (6.3, 6.9)$ and the Dirichlet boundary condition $\mathcal{B} = \mathcal{I}$ on $\Gamma = \{6.3, 6.9\}$, and follow the setting of operator reproducing kernel Hilbert space in Section~\ref{sec:simulation}. 
The results are presented in Table~\ref{tab:TX1}. 
As the tuning parameter $\lambda$ ranges form $10^{0.9}$ to $10^{1.9}$, we see that the $p$-value is always larger than $0.05$. 
This implies that the specification of differential operator $\mathcal{D}_1$ is retained at level $0.05$. 
Meanwhile, $\mathrm{RSS}_{\lambda} = \sum_{i=1}^{n}\lVert F_{i} - \hat{\mathcal{D}}_{\lambda}(U_{i})\rVert_{L^{2}(\Omega)}^{2}$ is around $277.1 < \mathrm{RSS}_{0}$, and $\mathrm{GCV}_{\lambda}$ is around $1.280$, which is slightly larger than $\mathrm{RSS}_{0}/n = 1.247$. 
Consequently, the proposed nonparametric estimator serves as a comparable surrogate. 

\begin{table}[!ht]
	\caption{Results of estimation and goodness-of-fit test for the differential equation \eqref{eq:thermo} using ERA5 data.}
	\label{tab:TX1}
	\centering
	\begin{tabular}{ccccccc}
		\hline
		$\lambda$ & $10^{0.9}$ & $10^{1.1}$ & $10^{1.3}$ & $10^{1.5}$ & $10^{1.7}$ & $10^{1.9}$ \\
		$\mathrm{RSS}_{\lambda}$ & 277.1344 & 277.1345 & 277.1348 & 277.1357 & 277.1378 & 277.1432 \\
		$\mathrm{GCV}_{\lambda}$ & 1.280014 & 1.280012 & 1.280011 & 1.280010 & 1.280012 & 1.280024 \\
		$p$-value & 0.271 & 0.245 & 0.251 & 0.235 & 0.241 & 0.242 \\
		\hline
	\end{tabular}
\end{table}


%
%


\bibliographystyle{agsm}
\bibliography{ref}
\end{document}